\RequirePackage{fix-cm}
\documentclass[smallextended]{svjour3}       % onecolumn (second format)
\smartqed  % flush right qed marks, e.g. at end of proof
\usepackage{subfigure}
\usepackage{bm,amsmath}
\usepackage{algorithm}
\usepackage{algpseudocode}
\usepackage{amsfonts}
\usepackage{graphicx}
\usepackage{enumitem}
\usepackage{mathrsfs}

\DeclareMathAlphabet\mathpzc{OT1}{pzc}{m}{it}
\let\mathcal=\mathpzc

\numberwithin{equation}{section}

\let\trueiiint=\iiint
\def\iiint{\mathop{\textstyle\trueiiint}\limits}
\def\intinfty{\int\limits_{\!\!-\infty\,\,}^{\,\,\infty\!\!}\kern-0.0em}
\def\iintinfty{\mathop{\int\!\!\int}\limits_{\!\!-\infty\,\,}^{\,\,\infty\!\!}\kern-0.0em}
\def\iiintinfty{\mathop{\int\!\!\int\!\!\int}\limits_{\!\!-\infty\,\,}^{\,\,\infty\!\!}\kern-0.0em}

\let\<=\langle
\let\>=\rangle
\let\^=\hat
\def\~#1{{\-ox{\sf#1}}}

\let\-=\mathbf

\let\-=\mathbf

\usepackage{algorithm}
\usepackage{algpseudocode}

\newtheorem{condition}[theorem]{Condition}

\def\@#1{{\cal #1}}

\begin{document}

\title{Bayesian inverse regression for dimension reduction with small datasets\thanks{This work was
supported by the NSFC under grant number 11301337. The computational resources were
provided by the Student Innovation Center as Shanghai Jiao Tong University.}}


\titlerunning{Bayesian inverse regression for dimension reduction}        % if too long for running head

\author{Xin Cai         \and
        Guang Lin \and
				Jinglai Li %etc.
}

\authorrunning{X. Cai, G. Lin and J. Li} % if too long for running head

\institute{X. Cai \at
              School of Mathematical Sciences
Shanghai Jiao Tong University, 800 Dongchuan Rd, Shanghai 200240, China.
\and G. Lin \at
Department of Mathematics, Purdue University, West Lafayette, IN 47906, USA.
           \and  J. Li \at
              Department of Mathematical Sciences, University of Liverpool, Liverpool L69 7XL, UK.\\
\email{Jinglai.Li@liverpool.ac.uk}
}

\date{Received: date / Accepted: date}

\maketitle

\begin{abstract}
We consider supervised dimension reduction problems, namely to identify a low dimensional projection
of the predictors $\-x$ which can retain the statistical relationship between $\-x$ and the response variable $y$.
We follow the idea of the sliced inverse regression (SIR) and the
sliced average variance estimation (SAVE) type of methods, which is to use the statistical information of
the conditional distribution $\pi(\-x|y)$ to identify the dimension reduction (DR) space.
In particular we focus on the task of computing this conditional distribution without slicing the data. We propose a Bayesian framework to compute the conditional distribution
where the likelihood function is obtained using the Gaussian process regression model. The conditional distribution $\pi(\-x|y)$ can then be computed
directly via Monte Carlo sampling. We then can perform DR by considering
certain moment functions (e.g. the first or the second moment) of the samples of the posterior distribution.
With numerical examples, we demonstrate that the proposed method is especially effective for small data problems.

\keywords{Bayesian inference \and
covariance operator \and
dimension reduction \and
Gaussian process\and
inverse regression.
}
 \subclass{62F15 \and 65C05}
\end{abstract}

  \section{Introduction}\label{sec:intro}

In many statistical regression problems, one has
to deal with problems where the available data are insufficient to provide a robust regression.
If conducting regression directly in such problems, one often risks of overfitting or
being incorrectly regularized. In either case, the resulting regression model
may lose its prediction accuracy.
Extracting and selecting the important features or eliminating the redundant ones is a key step
to avoid overfitting and improve the robustness of the regression task~\cite{fukumizu2004dimensionality}.
The feature extraction and selection thus constitutes of identifying a low dimensional subspace of the predictors $\-x$ which
retains the statistical relationship between $\-x$ and the response $y$, i.e. a supervised dimension reduction problem.
Mathematically such problems are often posed as to estimate the central dimension reduction~(DR) subspace~\cite{cook2005sufficient}.
A very popular class of methods estimate this central subspace by considering the statistics of the predictors $\-x$ conditional on
the response $y$, and such methods include the sliced inverse regression (SIR) proposed in
the seminal work~\cite{li1991sliced}, the sliced average variance estimation~\cite{cook1991sliced,dennis2000save}, and many of their variants,
e.g. \cite{cook2005sufficient,li2007directional,zhu2007kernel,li2008sliced,xia2002adaptive,li2009dimension,ma2012semiparametric,tan2018convex,lin2019sparse}.
Some of the extensions and variants have been developed specifically for machine learning
problems, e.g., \cite{fukumizu2012gradient,wu2009localized,kim2008dimensionality}.
%Theoretical analysis of these methods, especially their asymptotic property, has also been extensively %studied~\cite{zhu1995asymptotics,hsing1992asymptotic,zhu1996asymptotics}.
The literature in this topic is vast and we refer to \cite{ma2013review,li2018sufficient} for a more comprehensive overview
of the subject.
It should be noted that most of the aforementioned methods adopt nonparametric formulation without assuming any specific relation between $\-x$ and $y$. 
As will be shown in the examples, the nonparametric approaches may not work well for
the problems with very small number of data, which is considered in the present work. 
To this end, an alternative type of methods is to assume a parametric model of the likelihood function $p(y|\-x)$,
and then compute the reduced dimensions with maximum likelihood estimation~\cite{cook2011ldr,cook2009likelihood}, or in a Bayesian formulation~\cite{mao2010supervised,reich2011sufficient}.
A main disadvantage of the parametric models is that they may be lack of the flexibility to 
accurately characterize of the relation between the predictors $\-x$ and the response $y$.

In this work we present a  method  incorporating the SIR/SAVE type of methods with the model base ones, 
to make them more effective for small data problems.  
In particular we remain in the SIR/SAVE framework to identify the DR space.
As one can see,  many  works in this class focus on the question:
what statistical information of the conditional distribution $\pi(\-x|y)$ should one use to obtain the DR subspace?
For example, SIR makes use of the expectation of $\pi(\-x|y)$ to identify the DR directions, SAVE utilizes
the variance of it, and the method in \cite{yin2003estimating} is based on the third moments.
In this work  we consider a different aspect of the problem: how to obtain the conditional distribution $\pi(\-x|y)$
 when the data set is small?
In SIR and SAVE, the conditional moments are approximately estimated by slicing the data~\cite{li1991sliced}.
As will be demonstrated with numerical examples, the slicing strategy does not perform well if
we have a very small data set.
The main purpose of the work is to address the problem of computing the conditional distribution $\pi(\-x|y)$.
In particular we present a Bayesian formulation which can provide not only the first or second moments,
but the full conditional distribution $\pi(\-x|y)$, and once the conditional distribution is available one can use any desired methods to estimate the DR subspace based on the conditional distribution.
Just like \cite{cook2011ldr,cook2009likelihood}, our method also involves constructing the likelihood function $\pi(y|\-x)$ from data, but a main difference here is that
 we characterize the likelihood function with a nonparametric Gaussian Process (GP) model~\cite{williams2006gaussian}, 
 which may provide more flexibility than a parametric model.
Once the likelihood function is available, we can compute the posterior distribution $\pi(\-x|y)$ from the likelihood function
and a desired  distribution of $\-x$.
In this work we  choose to mainly use the first order moment of the conditional distribution (following SIR) to demonstrate the method, while noting that the method can be easily extended to other conditional moments.
It is important to note here that, while the conditional distribution $\pi(\-x|y)$ is computed in a Bayesian fashion,
the core of the method (i.e. the estimation of the DR subspace) remains frequentist, 
and so it is fundamentally different from the
 methods~\cite{tokdar2010bayesian,mao2010supervised,reich2011sufficient} that do estimate 
the DR subspace with a Bayesian formulation (e.g., imposing a prior on the DR subspace). 

%These methods, however, are fundamentally different from the present one.
%First in all these existing methods, prior distributions need to be assumed on the DR subspace or the related parameters,
%while in the proposed method, we do not impose any prior assumptions on the DR subspace.
%Second in \cite{mao2010supervised,reich2011sufficient}, either the forward~\cite{reich2011sufficient}
%or the inverse~\cite{mao2010supervised} conditional distribution is assumed to be in a mixture form,
%while our method uses a GP nonparametric framework to model the conditional distribution $\pi(y|\-x)$,
%which can provide more flexibility than a mixture model.
%Finally in all the model-based approaches, Bayesian or non-Bayesian,
%the DR directions are either sampled with Markov chain Monte Carlo (MCMC) or
%computed with Maximum likelihood estimation (MLE),
%and both approaches can be rather computationally intensive.
%The proposed method here does not need to perform either MCMC or MLE,
%and its computational cost is about the same as the standard SIR.

%We summarize three main advantages of the proposed Bayesian method:
To summarize, the main contribution of the work is to propose a GP based Bayesian formulation to compute the conditional distribution 
$\pi(\-x|y)$ for any value of $y$ in the SIR/SAVE framework, and by doing so it avoids slicing the samples,
which makes it particularly effective for problems with very small numbers of data.
%The method uses non-parametric GP model to construct the likelihood function, which can
%rather flexibly  and efficiently characterize the relation between $\-x$ and $y$.
%Finally, the method does not require MCMC simulation or optimization~(such as MLE),
%which makes it computationally competitive against the existing model based methods.

The rest of this paper is organized as follows. In section \ref{sec:sliced} we set up a formulation of dimension reduction and go through the basic idea of the classic dimension reduction approaches SIR and SAVE. The Bayesian inverse regression and 
the Bayesian average variance estimation are introduced explicitly in Section \ref{sec:bir}, including a Bayesian formulation
for computing $\pi(\-x|y)$, the GP model used, and complete algorithms to draw samples from $\pi(\-x|y)$. In section \ref{sec:examples} we provide several numerical examples. Section \ref{sec:conclusion} offers concluding remarks.

\section{Dimension reduction and the sliced methods}\label{sec:sliced}
\subsection{Problem setup}

We consider a generic supervised dimension reduction problem.	Let $\-x$ be a $p$-dimensional random variable defined on $R^p$ following a distribution $\pi_0(\-x)$,
and suppose that we are interested in a scalar function of $\-x$, which ideally can be written as,
	\begin{equation}
	{y}=f(\-b_1^T\-x, \-b_2^T\-x,..., \-b^T_K\-x,\epsilon), \label{1}
	\end{equation}
	 where $\-b_k$ for $k=1...K$ are some $p$-dimensional vectors,  and $\epsilon$ is small noise independent of $\-x$.
	 It should be clear that, when this model holds, the projection of the $p$-dimension variable $\-x$  onto the $k$ dimensional
	subspace of $R^p$ spanned by $\{\-b_1,...,\-b_K\}$, captures all the information of $\-x$ with respect to $y$, and if $K<p$, we can achieve the goal of data reduction by estimating the coefficients $\{\-b_k\}_{k=1}^K$.
	 In practice, both the explicit expression of $f$ and the coefficients $\{\-b_k\}_{k=1}^K$ are unknown, and instead we
	 have a set of data pairs $\{(\-x_j,y_j)\}_{j=1}^n$ drawn from the joint distribution $\pi(\-x,y)$ defined by $\pi_0$ and Eq.~\eqref{1}.
	 Finding a set of $\{\-b_k\}_{k=1}^K$ that satisfy the Eq.~\eqref{1} from the given data set
	 $\{(\-x_j,y_j)\}_{j=1}^n$ is the task of supervised dimension reduction.
	In what follows we shall refer to the
	coefficients $\{\-b_k\}_{k=1}^K$ as  dimension-reduction (DR) directions, and the linear space $B$ spanned by the $\{\-b_k\}_{k=1}^K$ as the DR subspace.
For a more formal and generic description of the DR problem~(in the Central DR Subspace and Sufficient Dimension Reduction framework)  we refer to \cite{cook2005sufficient}.

\subsection{Sliced inverse regression}

The SIR approach~\cite{li1991sliced} estimates the DR directions based on the idea of inverse regression (IR). In contrast to the forward regression
		$E({y}\,|\, \-x)$, IR regresses each coordinate of $\-x$ against $y$. Thus as $y$ varies, $E(\-x\,|\,{y})$ draws a curve in $R^p$ along the
		${y}$ coordinate, whose center is located at $E(E(\-x\,|\,{y}))=E(\-x)$.
		For simplicity we shall assume that throughout this section $\-x$ is a standardized random variable:
		namely $E(\-x)=0$ and $\mathrm{Cov}(\-x)= I$.
		%under the model in Eq.~\eqref{1}, this curve will hover around a K-dimensional affine subspace with the following condition:	
		Under the following condition the IR curve $E(\-x\,|\,y)$ is contained in the DR subspace $B$~\cite{li1991sliced}:
	\begin{condition} For any $\beta\in R^p$, the conditional expectation $E(\beta^T\-x\,|\,\-b^T_1\-x,...,\-b^T_K\-x)$ is linear in
	$\-b^T_1\-x,..., \-b_K^T\-x$. \label{cond:1}
	\end{condition}
	This condition is satisfied when the distribution of $\-x$ is elliptically symmetric~\cite{li1991sliced}.
	An important implication of this property is that the covariance matrix $\mathrm{Cov}[E(\-x\,|\,y)]$ is degenerated in any direction orthogonal to the DR subspace $B$. We see, therefore, that the eigenvectors associated with the largest $K$ eigenvalues of $\mbox{Cov}[E(\-x\,|\,y)]$ are the DR directions. So the key of estimating the DR direction is to obtain the covariance of the conditional expectation of the data,
	$\mbox{Cov}[E(\-x\,|\,{y})]$.
%	which then require the knowledge of $E(\-x|y)$.

One of the most popular approaches to estimate the covariance $\mbox{Cov}[E(\-x|y_j)]$ is SIR. Simply put, SIR
 produces a crude estimate of $E(\-x|y)$,  by slicing the data ${(\-x_1,y_1),...,(\-x_n,y_n)}$ into $H$ partitions according to the value
of ${y}_j$ and then estimating $E(\-x\,|\,y\in I_h),h=1,...,H$
	 using the data inside the interval $I_h$ for each $h=1,...,H$.
	  Finally one use the $H$ samples to compute an estimate of the covariance matrix $\mbox{Cov}[E(\-x|y)]$.
		A complete SIR scheme is described as follows:
	\begin{enumerate}
	\item Divide range of $y$ into $H$ slices, $I_1,...,I_H$.
	Let the proportion of the ${y}_j$ that falls in slice $I_h$ be $\hat{p}_h$, i.e.,
	\[\hat{p}_h = \frac1n\sum_{j=1}^n\delta_h({y}_j),\]
	where $\delta_h(\text{y}_j)$ takes the values 0 or 1 depending on whether $y_j$ falls into the $h$th slice $I_h$ or not.
	\item Within each slice, compute the sample mean of the ${\-x}_j$'s, denoted by $\hat{\-x}_h(h=1,...,H)$:
	\[\hat{\-x}_h = \frac1{(n\hat{p}_h)}\sum_{\text{y}_j\in I_h}{\-x_j}.\]
   \item %Conduct a weighted principal component analysis for the data $\{\hat{\-x}_h\}_{h=1}^H$:
	Compute the weighted covariance matrix
	\[\hat{C}=\sum_{h=1}^{H}\hat{p}_h\hat{\-x}_h\hat{\-x}_h^T.\]
%	then find the eigenvalues and the eigenvectors for $\hat{V}$.
    \item Perform eigenvalue decomposition of $\hat{C}$, and return
	the eigenvectors associated with the $k$ largest eigenvectors as the estimated DR directions $\hat{\-b}_1,...,
	\hat{\-b}_K$.
		\end{enumerate}
		
		As is mentioned in Section~\ref{sec:intro}, the slicing treatment is often not sufficiently accurate when the data set is small,
	and in what follows	we shall provide an alternative  to compute the covariance matrix.

	\subsection{Sliced average variance estimation}
        The SAVE method extract the DR directions from the variance of $\pi(\-x|y)$,
        %There may be information available in the inverse variance function as well. Consider the case, $\-x = (\-x_1,\-x_2)\sim \mathcal{N}(0,I_2), y = \-x_2^2 + \epsilon, \epsilon \sim \mathcal{N}(0,1)$ and $\epsilon $ independent of $\-x$. Apparently $\{(0,1)^T\}$ spans the DR space,
  %  while $E(\-x|y) = 0$ and SIR doesn't work out. 
and by doing so it is able to recover the information that could be overlooked by SIR because of symmetries in the forward regression 
function~\cite{dennis2000save}. Let the columns of $\-b$ form a basis for the DR space. To use SAVE, we need to assume the following two conditions~\cite{dennis2000save}:
        \begin{enumerate}
            \item $E(\-x|\-B^T\-x) ]$ is linear in $\-B^T\-x$,
            \item $\mbox{Var}(\-x|\-B^T\-x) $ is a constant,
        \end{enumerate}
        where $\-B$ is any basis matrix of $R^p$.
        The conditions hold when $\-x$ is normally distributed although normality is not necessary.
%where $P_\-b$ is a projection operator for the DR space, 
%$Q_\-b = I_p - P_\-b$, and $I_p$ is the identity matrix of order $p$. 
%Then
  %  \begin{equation}
   % \Sigma_{\-x|y} = Q_\-b + P_\-b\Sigma_{\-x|y}P_\-b
   % \end{equation}
  Under these two conditions, one can derive that  
	$$\mbox{span}\{I_p-E(\mbox{Cov}[{\-x|y}])\}$$ is a DR space~\cite{dennis2000save},
  which is the basis for SAVE.  A complete SAVE scheme is as follows:
    \begin{enumerate}
    \item Divide range of $y$ into $H$ slices, $I_1,...,I_H$.
	Let the proportion of the ${y}_j$ that falls in slice $I_h$ be $\hat{p}_h$, i.e.,
	\[\hat{p}_h = \frac1n\sum_{j=1}^n\delta_h({y}_j),\]
	where $\delta_h({y}_j)$ takes the values 0 or 1 depending on whether $y_j$ falls into the $h$th slice $I_h$ or not.
    \item Within each slice, compute the sample covariance matrix of the ${\-x}_j$'s, denoted by $\hat{M}_h(h=1,...,H)$:
          \begin{equation}
          \hat{M}_h = \sum_{\text{y}_j\in I_h}{\-x_j\-x_j'}.
          \end{equation}
    \item The $j$-th sample SAVE DR direction can now be constructed by perform eigenvalue decomposition on the following matrix , and return the eigenvectiors associated with the $k$ largest eigenvectors:
           \begin{equation}
           \hat{C} = \sum_{h=1}^{H}\hat{p}_h{(I-\hat{M}_h)^2}
           \end{equation}
    \end{enumerate}

\section{Bayesian inverse regression}\label{sec:bir}

\subsection{Bayesian formulation for $\pi(\-x|y)$}
Recall that in the SIR framework, a key step is to compute the covariance $\mbox{Cov}[E(\-x|y)]$.
A natural choice to estimate the covariance $\mbox{Cov}[E(\-x\,|\,{y})]$ is to use the sample covariance of the
data points,
\begin{equation}
\hat{C}= \frac1{n-1} \sum_{j=1}^n (\hat{\-x}_j - \bar{\-x})(\hat{\-x}_j - \bar{\-x})^T,
\quad \bar{\-x} = \frac1n\sum_{j=1}^n \hat{\-x}_j, \label{e:cov}
\end{equation}
where $\hat{\-x}_j$ is an estimate of $E(\-x|y_j)$ for all $j=1...n$,
and $(y_1,...,y_n)$ are the data points.
Next we need to compute   $\hat{\-x}_j$, the estimate of $E(\-x|y_j)$, and we
propose to do so in a Bayesian framework.
Namely we formulate the problem as to compute the posterior distribution:
\begin{equation}
\pi(\-x|y) \propto {\pi(y|\-x)\pi(\-x)},\label{e:postdis}
\end{equation}
where $\pi(y|\-x)$ is the likelihood function and $\pi(\-x)$ is the prior of $\-x$.

We consider the prior distribution $\pi(\-x)$ first.
%Next we shall determine the prior distribution $\pi(\-x)$.
To start we note that the choice of prior does not affect the DR subspace as this subspace structure
lies in the function $f(\-x,\epsilon)$ in Eq.~\eqref{1} rather than the distribution of $\-x$.
As such, in principle one may use any prior distribution that satisfies the conditions required by SIR/SAVE.
% rather than the original distribution $\pi_0(\-x)$.
However,  if the chosen $\pi(\-x)$ is too different from $\pi_0$, the GP model constructed from the data (following $\pi_0(\-x)$)
may not be accurate for the samples drawn according to $\pi(\-x)$, which in turn may hurt the accuracy of the posterior
$\pi(\-x|y)$.
To this end, one should choose the prior to be $\pi_0$ or close to it. 
We consider the following three cases.
First in certain problems, especially those where the data are generated from computer models,
the distribution $\pi_0(\-x)$ may be known in advance.  
Secondly for most problems where $\pi_0$ is not available in advance, a natural choice is
to perform a crude density estimation for the data $\{\-x_j\}_{j=1}^n$ and use the estimated density as the prior.
For example, one may use Gaussian mixtures~\cite{mclachlan2004finite} or a simple Gaussian to estimate the prior distribution from the data $\{\-x_j\}_{j=1}^n$.
Finally, for problems where estimating the density of $\-x$ are particularly challenging, 
 we can just use the original data points
 $\{\-x_j\}_{j=1}^n$, and in this case the prior is simply $\pi_0$.
%In what follows we shall use the latter strategy, i.e., to use the original data points,
%for reasons that will be discussed later.

\subsection{The GP regression}
The next step is to construct the likelihood function $\pi(y|\-x)$ from data,
which, as mentioned earlier, is done by using the GP regression model.

Simply speaking the GP regression performs a nonparametric regression in a Bayesian framework~\cite{williams2006gaussian}.
%and more precisely we want to compute the conditional distribution $\pi(y|\-x,D)$.
%The main idea of the GP method is to assume that the data points and the new point $(\-x,y)$ are
%from a Gaussian random field defined on $R^{n_x}$,
The main idea of the GP method is to assumes that the function of interest $f(\-x,\epsilon)$ is a realization from a Gaussian random field,
whose mean is $\mu(\-x)$ and covariance is specified by
a kernel function $k(\-x,\-x')$, namely,
\[ \mathrm{Cov}[f(\-x),f(\-x')] = k(\-x,\-x'). \]
The kernel $k(\-x,\-x')$ is positive semidefinite and bounded.
%whose mean is $\mu(\-x)$ and covariance is given by a  kernel function:
% \[ \mathrm{Cov}[y(\-x),y(\-x')]=k(\-x,\-x').\]

Now given the data points $\{(\-x_j,y_j)\}_{j=1}^n$,
%where $\{\-x_j\}_{j=1}^n$ are drawn from $\pi(\-x)$, and,
% \[{y}^*_i = f(\-x_i^*)\quad \mathrm{for} \quad i=1,\ldots,m.\]
we want to predict the value of $y$ at a new point $\-x$.
Now we let $\-X := \left[\-x_1, \ldots, \-x_n\right]$,
and $\-Y =[y_1,\ldots, y_{n}]$.
%Now let us assume that $m$ evaluations of the function $g(\-x)$ are performed
% at parameter values $\-X^* := \left[\-x^*_1, \ldots \-x^*_m\right]$, yielding  function evaluations $\-y^* := \left[ {y}^*_1, \ldots {y}^*_m\right]$,
Under the GP assumption,  it is easy to see that the joint distribution of $(\-Y,\,y)$ is Gaussian,
\begin{equation}
\left[ \begin{array}{c}
         \-Y \\
        y \end{array} \right] \sim \@N\left(\begin{array}{c}
         \mu(\-X) \\
        \mu(\-x) \end{array},
				\left[
				\begin{array}{ll}
         K(\-X,\-X)+\sigma_n^2I &K(\-X,\-x) \\
        K(\-x,\-X) &K(\-x,\-x) \end{array}\right]
				\right) , \label{e:jointdis}
\end{equation}
where $\sigma_n^2$ is the variance of observation noise, $I$ is
an identity matrix, and the notation $K(\-A,\-B)$ denotes the matrix of the covariance evaluated at all pairs of points in set $\-A$ and in set $\-B$ using
the kernel function $k(\cdot,\cdot)$.

It follows immediately from Eq.~\eqref{e:jointdis} that  the conditional distribution $\pi_{GP}(y|\-x,\-X,\-Y)$  is also Gaussian:
\begin{subequations}
\label{e:gp}
  \begin{equation}
	\pi_{GP}(y|\-x,\-X,\-Y) =\mathcal{N}(\mu_\mathrm{pos}, \sigma^2_\mathrm{pos}),
	\end{equation}
where the posterior mean and variance are,
\begin{align*}
&\mu_\mathrm{pos}(\-x)=\mu(\-x)+k(\-x,\-X)(k(\-X,\-X)+\sigma_n^2I)^{-1}(\-Y-\mu(\-x)),\\
&\sigma^2_\mathrm{pos} = k(\-x,\-x)-k(\-x,\-X)(k(\-X,\-X)+\sigma_n^2I)^{-1}k(\-X,\-x).
\end{align*}
\end{subequations}
There are also a number of technical issues in the GP model, such as choosing the kernel function and determining the hyperparameters.
For detailed discussion of these matters, we refer the readers to~\cite{williams2006gaussian}.
In what follows we shall use the GP posterior as
the likelihood function, i.e., letting $\pi(y|\-x) =\pi_{GP}(y|\-x,\-X,\-Y)$.

%Second by choosing such a prior we ensure that the distribution of $\-x$ used to compute the conditional probability
%is elliptically symmetric.

\subsection{Computing the posterior mean}\label{sec:comppost}

%A main advantage of the latter approach is that it is significantly more efficient
%as it requires neither the density estimation procedure nor  the MCMC simulation
%which both can be computationally intensive.
%For this reason, in this work we choose to directly use the original data points.

Once we obtain the likelihood function and the prior, a straightforward idea is to draw samples
from the posterior distribution~\eqref{e:postdis} with the Markov chain Monte Carlo (MCMC) simulation.
An alternative strategy is to sample from $\pi(\-x)$ in an importance sampling (IS) formulation.
Namely suppose that we draw a set of samples $\{\-x_i\}_{i=1}^{n_{MC}}$ from the prior distribution $\pi(\-x)$,
and for each $\-x_i$ we can compute the weight $$w_i=\pi(y|\-x_i).$$
Finally the weights $w_1,...,w_{n_{MC}}$ are normalized so that
$\sum_{i=1}^{n_{MC}} w_i=1$ (if these samples are drawn
with MCMC, then $w_i=1/n_{MC}$ for all $i=1...{n_{MC}}$).
We thus obtain obtain a set of weighted samples
$\{(\-x_i,w_i)\}_{i=1}^{n_{MC}}$ drawn from the posterior $\pi(\-x|y)$.
Now let $\{(\-x_i,w_i)\}_{i=1}^{n_{MC}}$ be a set samples draw from the posterior, and
we can estimate $E(\-x|y)$ as
\begin{equation}
\hat{\-x}=\sum_{i=1}^{n_{MC}} w_i\-x_i.
\end{equation}
We repeat this procedure for each $y_j$ for $j=1...n$, and then use Eq.~\eqref{e:cov} to compute $\mbox{Cov}[E(\-x|y)]$.
Since we use a Bayesian method to estimate $E(\-x|y)$, we refer to proposed method as
Bayesian inverse regression (BIR).
Similarly the samples can also be used to estimate the conditional covariance $\mbox{Cov}[\-x|y]$ in SAVE,
and the resulting method is termed as Bayesian average variance estimation (BAVE). 
As is discussed earlier, the key of BIR/BAVE is  essentially provides a means to
draw samples from the conditional distribution $\pi(\-x|y)$ without slicing the data, and its application is not limited
to estimate $E(\-x|y)$ or $\mbox{Cov}[\-x|y]$, and it is possible to make use of the conditional distribution in a different manner.
Finally we present the BIR algorithm in Alg.~\ref{alg:bir}
and BAVE in Alg.~\ref{alg:bir2}.

\begin{algorithm}[t]
\caption{The Bayesian inverse regression algorithm with MCMC}\label{alg:bir}
\let\-=\mathbf
\begin{algorithmic}[1]
\Require{$\{(\-x_j,y_j)\}_{j=1}^n$, $n_{MC}$, $\pi(\-x)$}
\Ensure{The estimated DR directions: $\hat{\-b}_1,\,...,\,\hat{\-b}_K$}
\State Construct the GP model from data $\{(\-x_j,y_j)\}_{j=1}^n$: $\pi_{GP}(y|\-x,\-X,\-Y)$;

\For{$j=1$ to $n$}
%\For{$i=1$ to $n$}
 %  \State Let $w_i$ = $\pi_{GP}(y_j|\-x_i,\-X,\-Y)$;
  %  \EndFor
		%			\State Renormalize $\{w_i\}_{i=1}^n$ so that $\sum_{i=1}^nw_i=1$;
		\State Draw $n_{MC}$ samples from $\pi_{GP}(y_j|\-x,\-X,\-Y)\pi(\-x)$: $\{\-x_i\}_{i=1}^{n_{MC}}$;
					\State Compute
$\hat{\-x}_j=\frac1{n_{MC}}\sum_{i=1}^{n_{MC}}\-x_i$;
\EndFor
\State Compute $\hat{C}$ using Eq.~\eqref{e:cov} and $\{\hat{\-x}_j\}_{j=1}^n$;
\State Perform eigenvalue decomposition of $\hat{C}$;
\State Return the eigenvectors associated with the $k$ largest eigenvalues as $\hat{\-b}_1,...,\hat{\-b}_K$.
\end{algorithmic}
\end{algorithm}

\begin{algorithm}[t]
\caption{The Bayesian average variance estimation algorithm with MCMC}\label{alg:bir2}
\let\-=\mathbf
\begin{algorithmic}[1]
\Require{$\{(\-x_j,y_j)\}_{j=1}^n$,$n_{MC}$,$\pi(\-x)$}
\Ensure{The estimated DR directions: $\hat{\-b}_1,\,...,\,\hat{\-b}_K$}
\State Construct the GP model from data $\{(\-x_j,y_j)\}_{j=1}^n$: $\pi_{GP}(y|\-x,\-X,\-Y)$;

\For{$j=1$ to $n$}
\State Draw $n_{MC}$ samples from $\pi_{GP}(y_j|\-x,\-X,\-Y)\pi(\-x)$: $\{\-x_i\}_{i=1}^{n_{MC}}$;
%\State Draw $n_{MC}$ samples from $\pi(\-x)$:$\{\-x_i\}_{i=1}^{n_{MC}}$;
%\For{$i=1$ to $n_{MC}$}
%   \State Let $w_i$ = $\pi_{GP}(y_j|\-x_i,\-X,\-Y)$;
%    \EndFor
%					\State Renormalize $\{w_i\}_{i=1}^n$ so that $\sum_{i=1}^nw_i=1$;
				\State Compute
$\hat{\-x}=\frac1{n_{MC}}\sum_{i=1}^{n_{MC}}\-x_i$;
					\State Compute
$\hat{M}_j=\frac1{n_{MC}-1}\sum_{i=1}^n (\-x_i-\hat{\-x}) (\-x_i-\hat{\-x})^T$;
\EndFor
\State Compute $\hat{C} = \frac{1}{n}\sum_{j=1}^{n}(I_p-\hat{M}_j)^2$;
\State Perform eigenvalue decomposition of $\hat{C}$;
\State Return the eigenvectors associated with the $k$ largest eigenvalues as $\hat{\-b}_1,...,\hat{\-b}_K$.
\end{algorithmic}
\end{algorithm}

\begin{remark}
It is important to reinstate here that, the BIR/BAVE methods only use
the Bayes' formula to compute the conditional distribution $\pi(\-x|y)$, and the DR methods themselves are
 \emph{frequentist}. 
\end{remark}

\begin{remark}
A key step in the proposed method is to construct the likelihood $\pi(y|\-x)$ with GP.
It is well known that GP may not perform well as a regression model for high dimensional problems. 
Nevertheless, as demonstrate by the examples, 
while it is unable to provide accurate regression results, the resulting GP model 
are often adequate for the dimension reduction purposes.  
Moreover, as is stated earlier, in this work we focus on problems 
with modestly high dimensionality (less than 100) and a very limited number of data (hundreds or less). 
%So here we want to emphasize again that the goal of this work is to address 
%problems with 
\end{remark}

\begin{remark}
Another issue that should be mentioned here is how to select the number of the reduced dimensions;
since BIR is also a method based on the eigenvalue decomposition of $\mbox{Cov}[E(\-x|y)]$,
the methods used in \cite{li1991sliced} and related works, e.g., \cite{ferre1998determining}, can be used directly here.
\end{remark}

\section{Numerical examples}
\label{sec:examples}

In this section we compare the performance of the proposed BIR/BAVE method with a number of common methods: SIR, SAVE, 
likelihood-based DR~(LDR)~\cite{cook2011ldr}, the Localized SIR (LSIR), 
in three mathematical and two real-data examples.
The first example uses data simulated from a mathematical function, with which we want to exam the scalability of the
methods with respect to the dimensionality of the problem.
The second one is also a mathematical example, and with this example we compare the performance of different methods
affected by the non-ellipticity of the distribution of $\-x$.
The third example is used specifically to compare the two second moment methods: SAVE and BAVE. 
Our last two examples are based on real data, in which we compare the performance of different methods in the small data situation.
In the GP model used in all the examples, we set the prior mean $\mu(\-x)=0$, and choose the Automatic Relevance Determination (ARD) squared
exponential kernel~\cite{williams2006gaussian}:
\begin{equation}
k(\-x,\-x')=\sigma^2_0\exp(-\frac12\sum_{i=1}^p\frac{(x_i-x'_i)^2}{\lambda_i^2}),
\end{equation}
 where the hyperparameters $\sigma_0$, $\lambda_1\,...,\lambda_d$, and the $\sigma_n$ are determined by
maximum likelihood estimation~\cite{williams2006gaussian}.
In all the examples except the one in Section~\ref{sec:rosenbrock}, the prior is obtained by fitting 
a Gaussian distribution to the data, while for the example in Section~\ref{sec:rosenbrock}, we assume that the distribution $\pi_0$ is known, which is used as the prior.
In addition, in all the examples,  
10000 MCMC samples are used to represent the conditional distribution $\pi(\-x|y)$ in 
the BIR and BAVE methods. 
%In the first two mathematical examples,  we use BIR Algorithm~\ref{alg:bir} with 10000 MCMC samples, and
%in the two real data examples, we use BIR Algorithm~\ref{alg:bir2}.

\begin{figure*}
  %\centering
 \subfigure[]{ \includegraphics[width=0.33\linewidth]{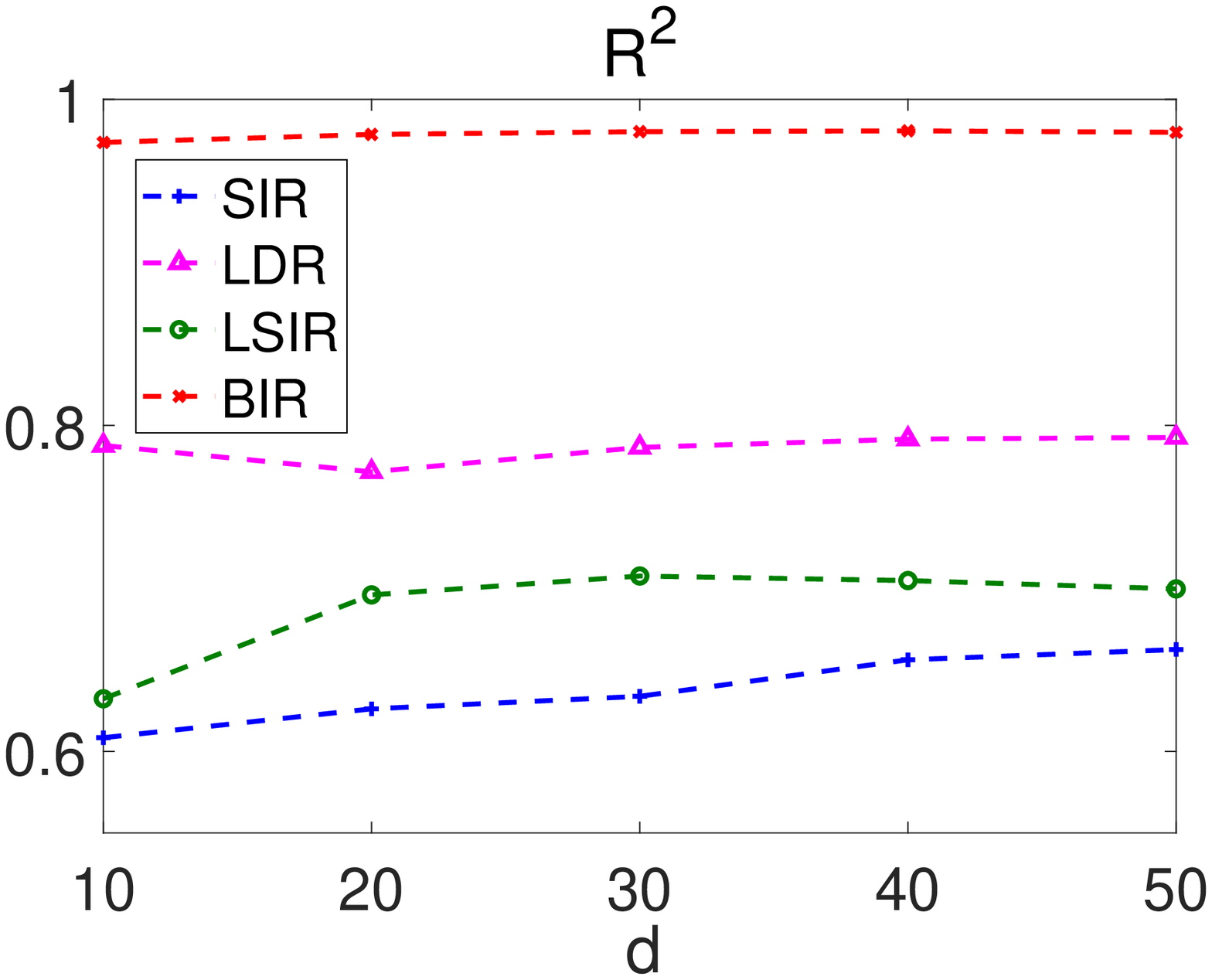}
	\includegraphics[width=0.33\linewidth]{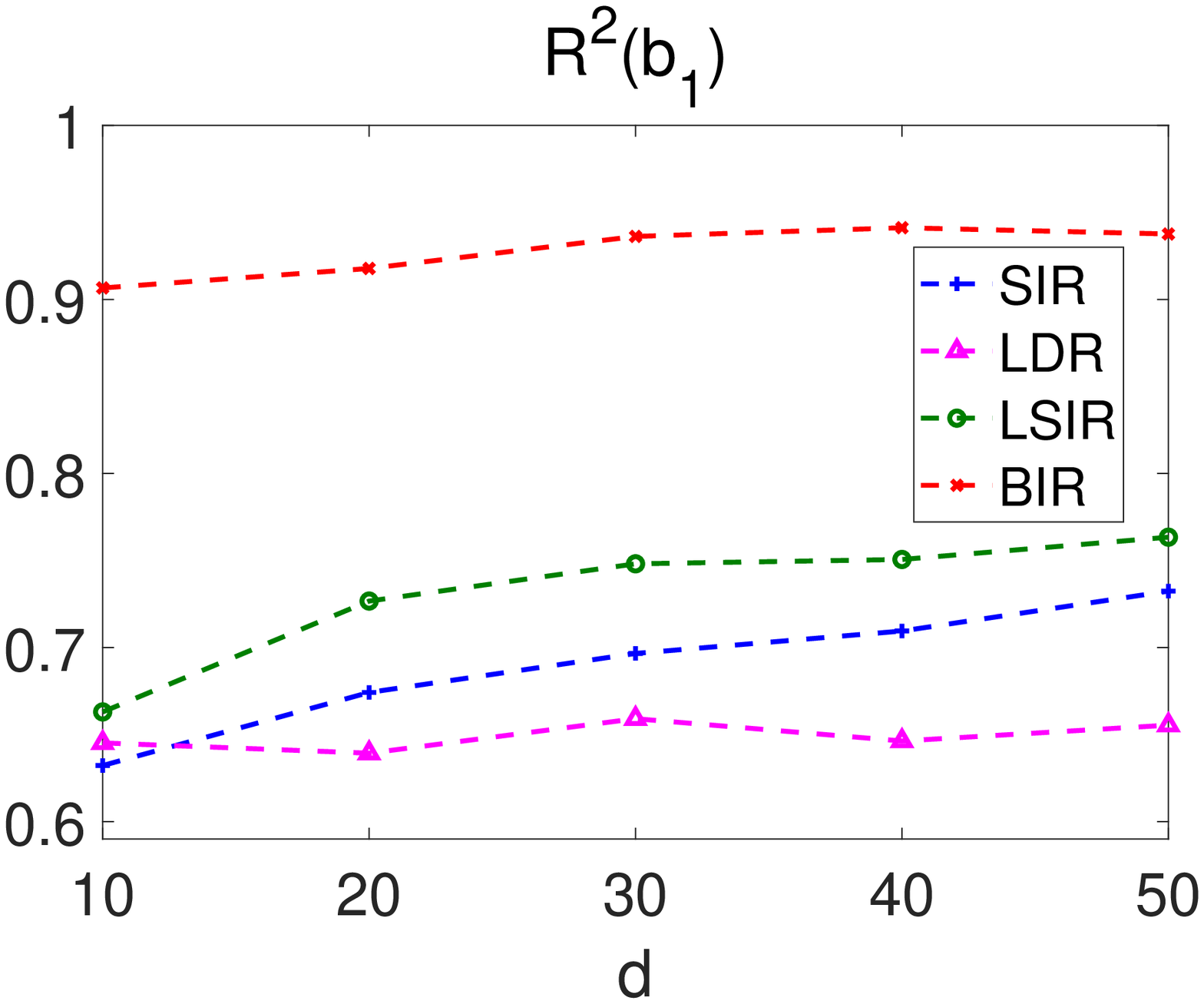}
  \includegraphics[width=0.33\linewidth]{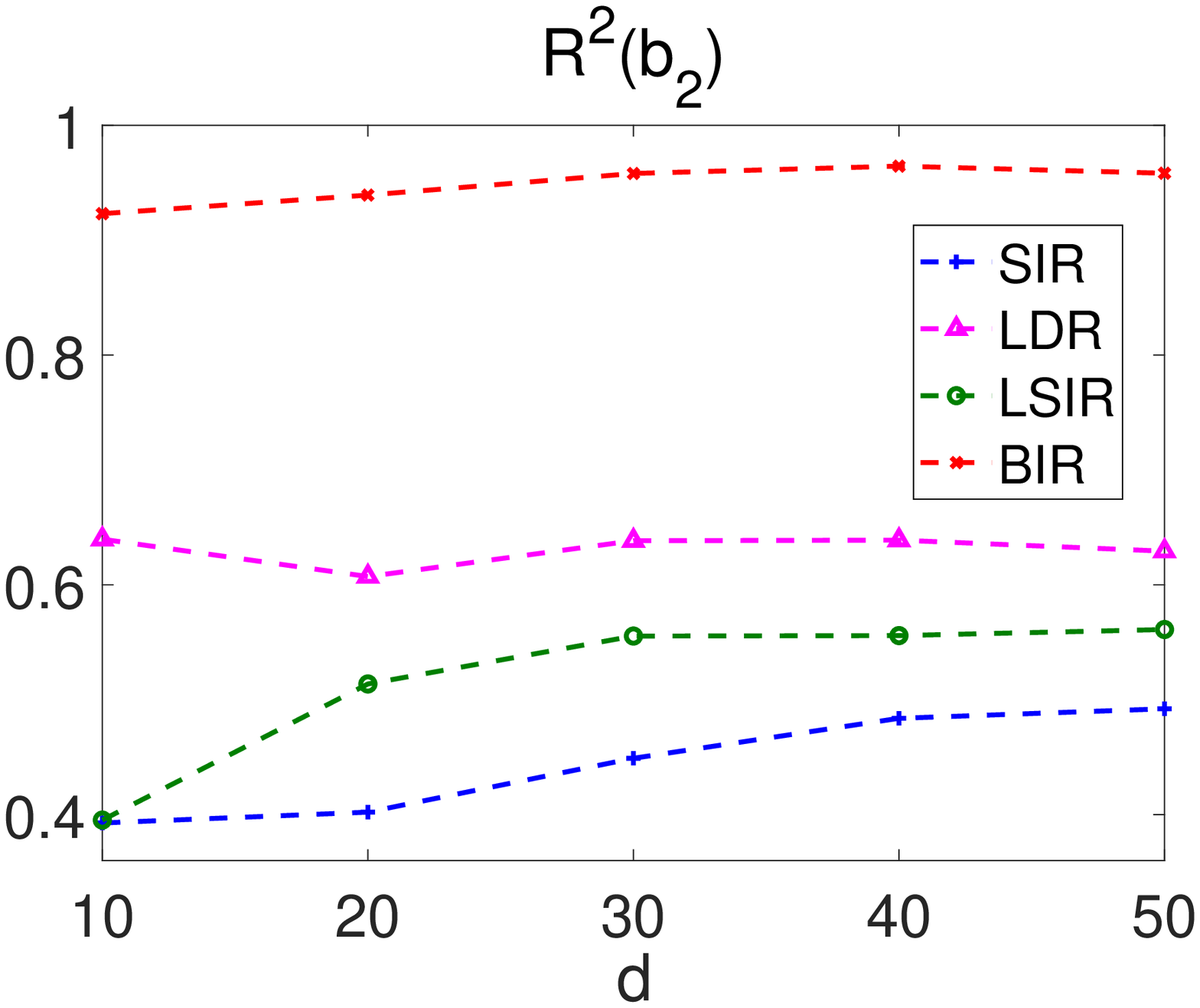}}
	%\caption{Results for function~\eqref{e:fun1}: the $R^2$ accuracy of the DR subspace (left), 
	 %the first DR direction $\-b_1$ (center) and the second DR direction
%$\-b_2$	(right),
	%all plotted against the dimensionality.}\label{f:eg11}
	\subfigure[]{\includegraphics[width=0.33\linewidth]{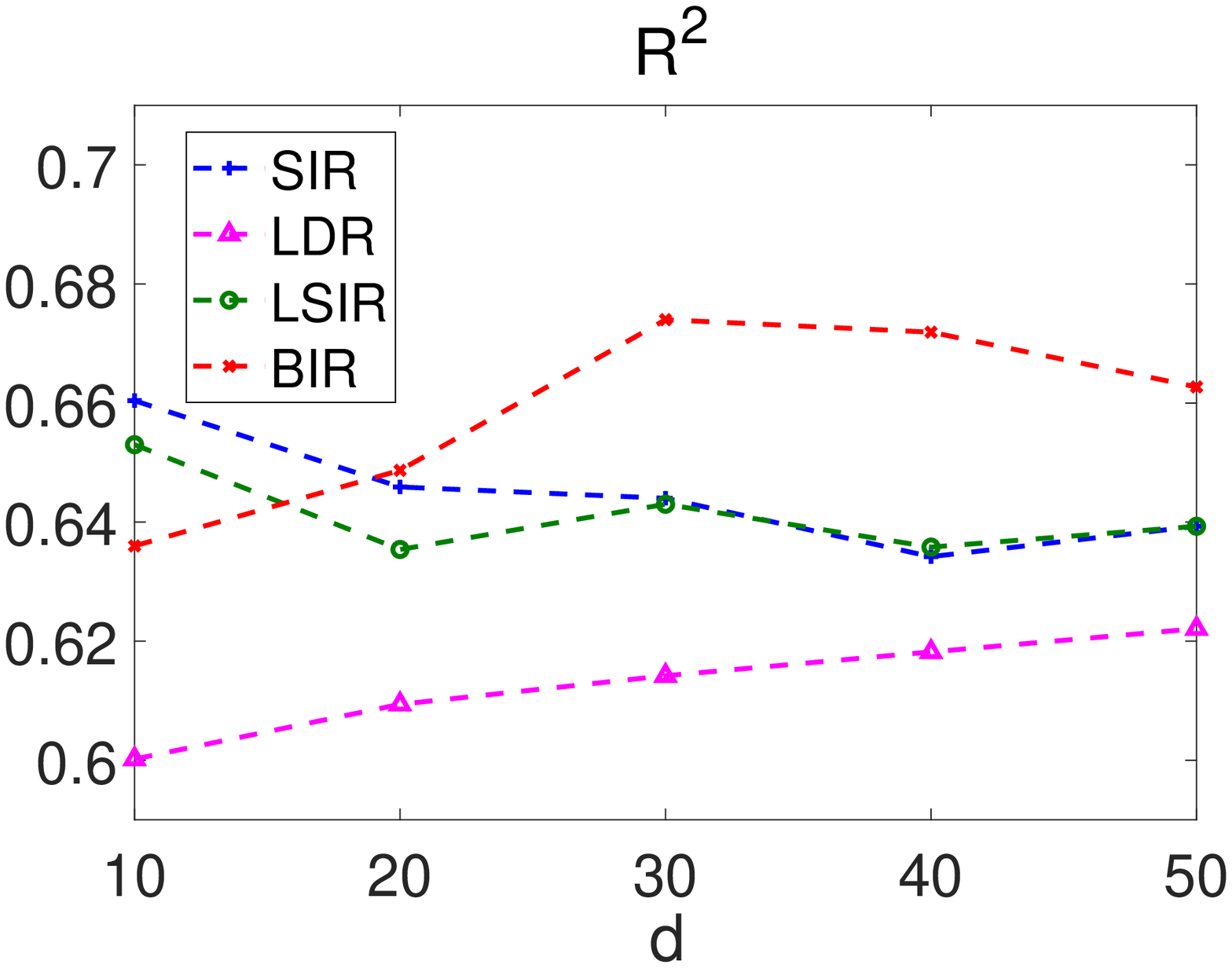}
	\includegraphics[width=0.33\linewidth]{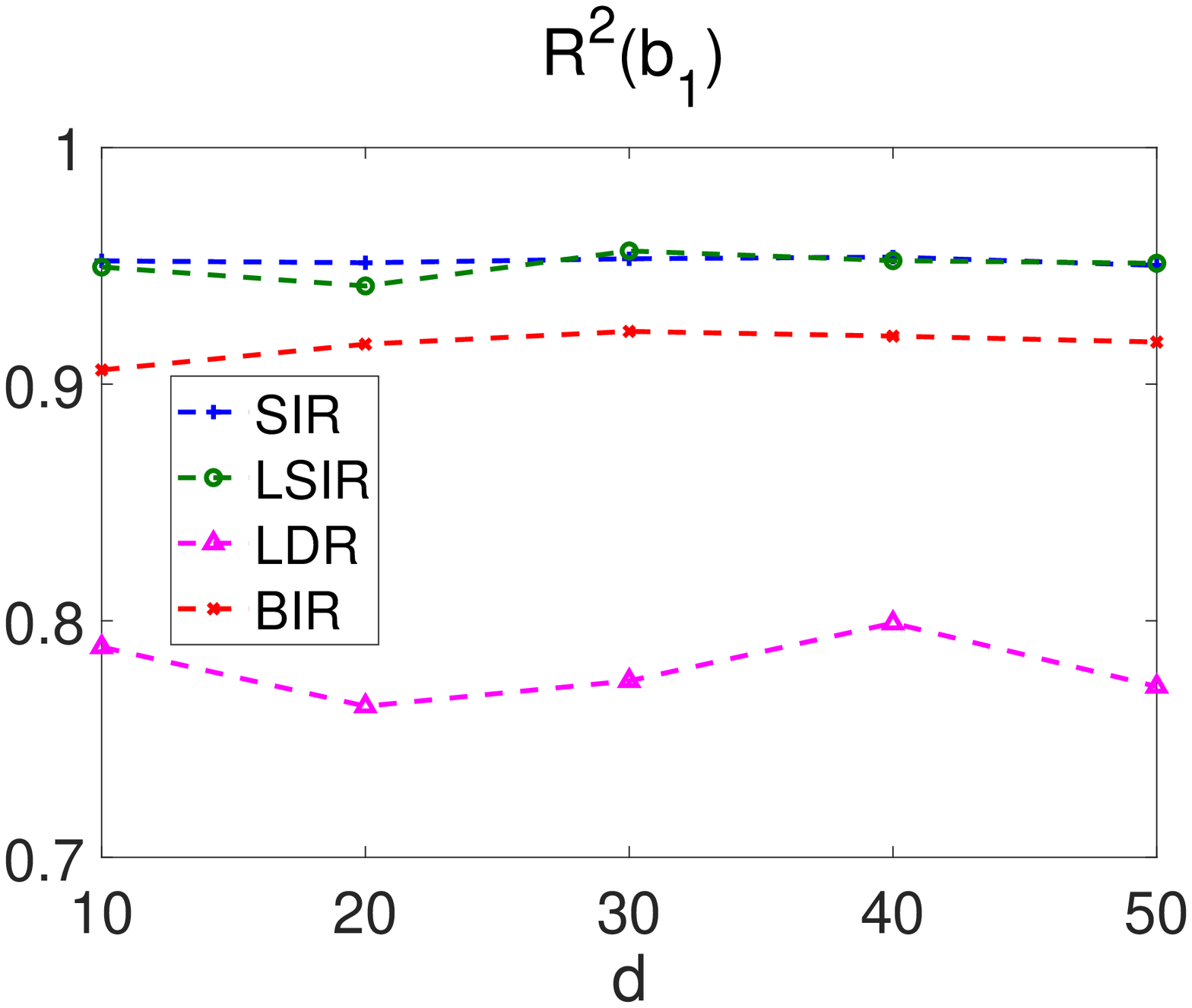}
  \includegraphics[width=0.33\linewidth]{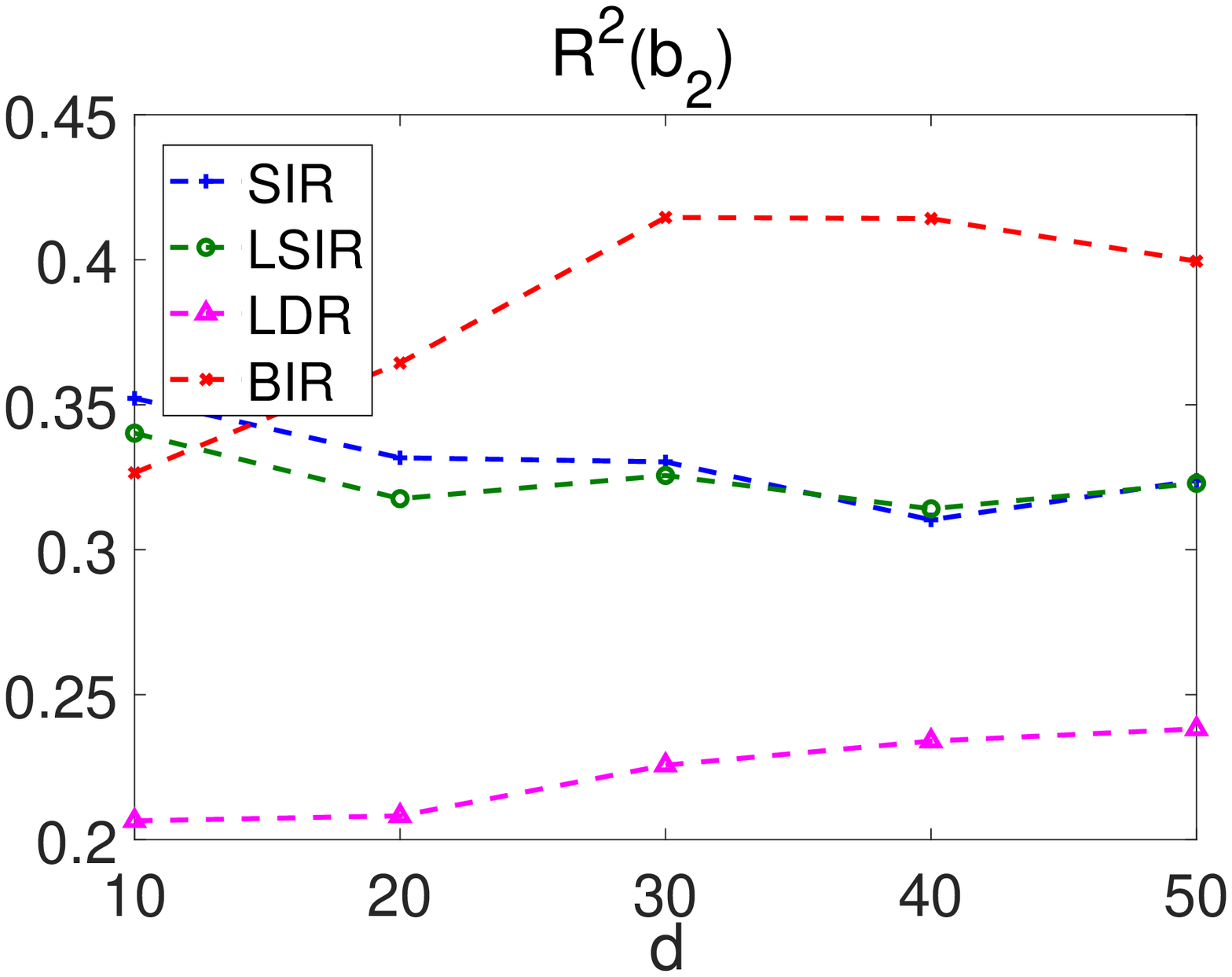}}
	\caption{The $R^2$ accuracy of the DR subspace (left), 
	 the first DR direction $\-b_1$ (center) and the second DR direction
$\-b_2$	(right), all plotted against the dimensionality.
(a) results for function~\eqref{e:fun1}; (b) results for function~\eqref{e:fun2}.} \label{f:eg1}
\end{figure*}

\subsection{Mathematical examples with increasing dimensions}
First we consider a $d$-dimensional problem where $\-x$ follows a standard normal distribution.
The data are simulated from the following functions:
\begin{subequations}
\begin{align}
\quad f(\-x,\epsilon) &= x_1(x_2+x_3)+0.5\epsilon,\label{e:fun1}\\
\quad f(\-x,\epsilon) &=\frac{x_1+x_2+x_3}{0.5+(x_4+x_5)^2}+0.1\epsilon,\label{e:fun2}
\end{align}
\end{subequations}
where $\epsilon\sim N(0,1)$.
Both problems have two DR directions.
In the regression content, a well known limitation of the GP method is that it can not handle high dimension, and so here we want to test the scalability of the BIR method with respect to dimensionality.
To do so we perform experiments for various dimensions: $d=10,\,20,\,30,\,40,\, 50$, where  we set the number of data points
to be $n=5d$, i.e., growing linear with respect to dimensionality.
To evaluate the performance of the methods, we use the $R^2$ metric of accuracy used in \cite{li1991sliced}
%\begin{equation}
 %R^2(b)=\max_{\beta\in B}\frac{(b\Sigma_{xx}\beta')^2}{b\Sigma_{xx}b'\cdot\beta\Sigma_{xx}\beta'}, \label{e:jointdis}
%\end{equation}
%the squared multiple correlation coefficient between the projected variable $b\-x$ and the ideally reduced variables $\beta_1\-x,\dots,\beta_K\-x$, and for a collection of $K$ estimated directions $b_1,\dots,b_K$ generating a linear subspace $\hat{B}$,
%\begin{equation}
 %R^2(\hat{B})=tr((B'B)^{-1}B'\hat{B}(\hat{B}'\hat{B})^{-1}\hat{B}'B), \label{e:jointdis}
%\end{equation}
to measure the accuracy of the DR subspace and the DR directions.

We repeat all the tests for $100$ times and report the average.
Specifically, we show the $R^2$-accuracy of the DR subspace $B$ and the two DR directions in
Figs.~\ref{f:eg1}. We can see that the BIR method has the best performance in all the tests in the two examples,
except one situation: $d=10$ for function~\eqref{e:fun2}.
The $R^2$ accuracy for each DR direction provide more information on the results.
Namely, for Function~\ref{e:fun1}, BIR performs better than all the other methods in both of the directions.
For function~\ref{e:fun2}, the accuracy of BIR is slightly lower than than SIR and LSIR for the first direction, but it achieves significantly higher accuracy on the second dimension than all the other ones.
Finally we want to note here that as the dimensionality increases, the performance of BIR does not decay evidently, suggesting
that the method can handle rather high dimensional problems.

\subsection{Mathematical examples with non Gaussian distributions}\label{sec:rosenbrock}
\begin{figure*}
  %\centering
 \centerline{ \includegraphics[width=1\linewidth]{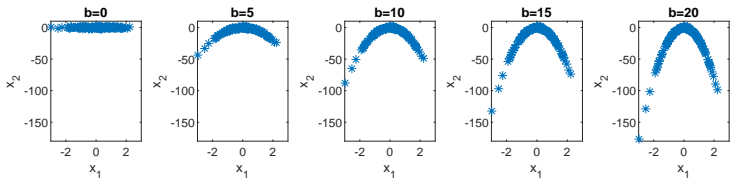}}
	\caption{The scatter plots of $(x_1,x_2)$ for different values of $b$.} \label{f:rosenbrock}
\end{figure*}

\begin{figure*}

\centerline{ \includegraphics[width=0.43\linewidth]{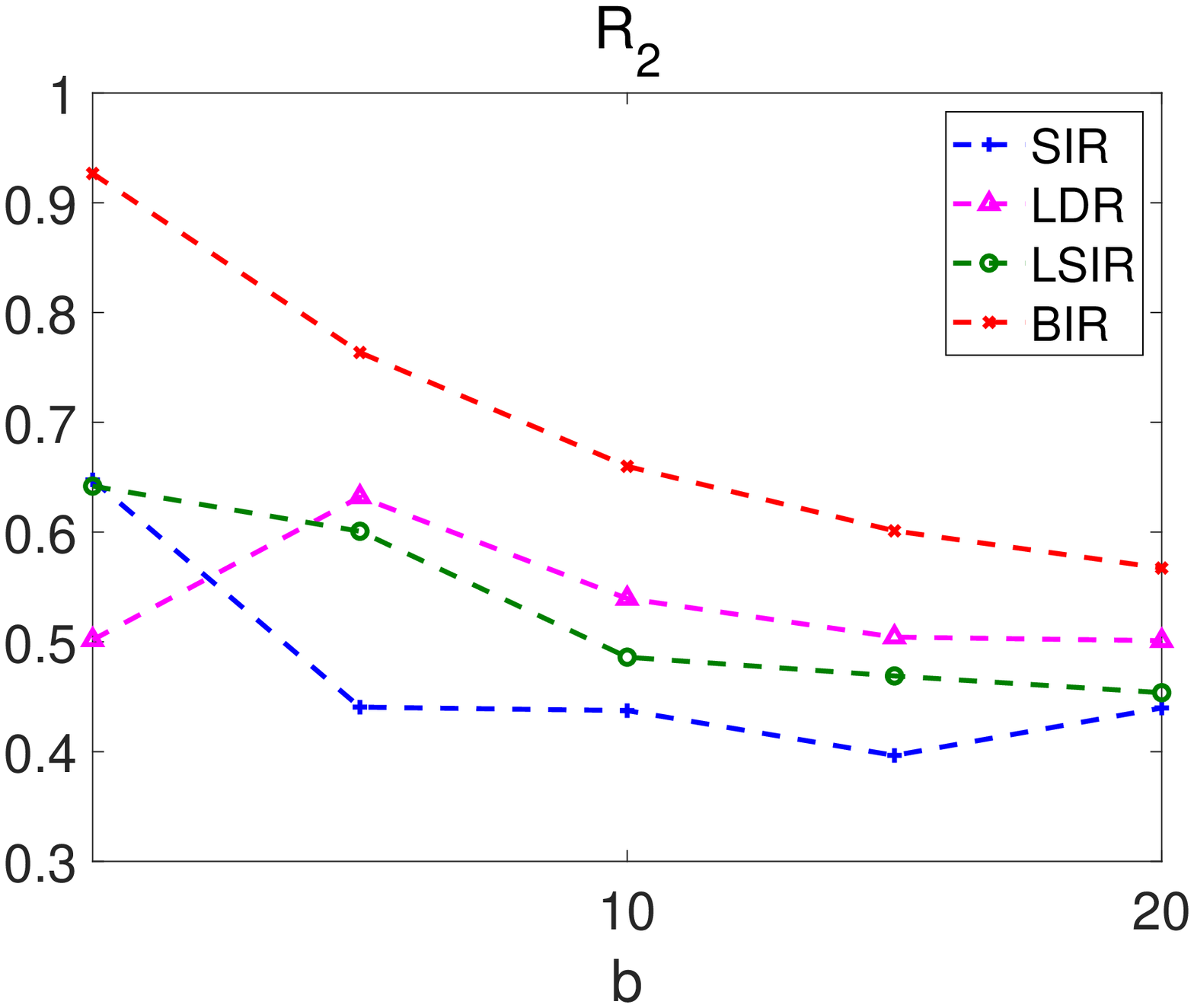}
	\includegraphics[width=0.43\linewidth]{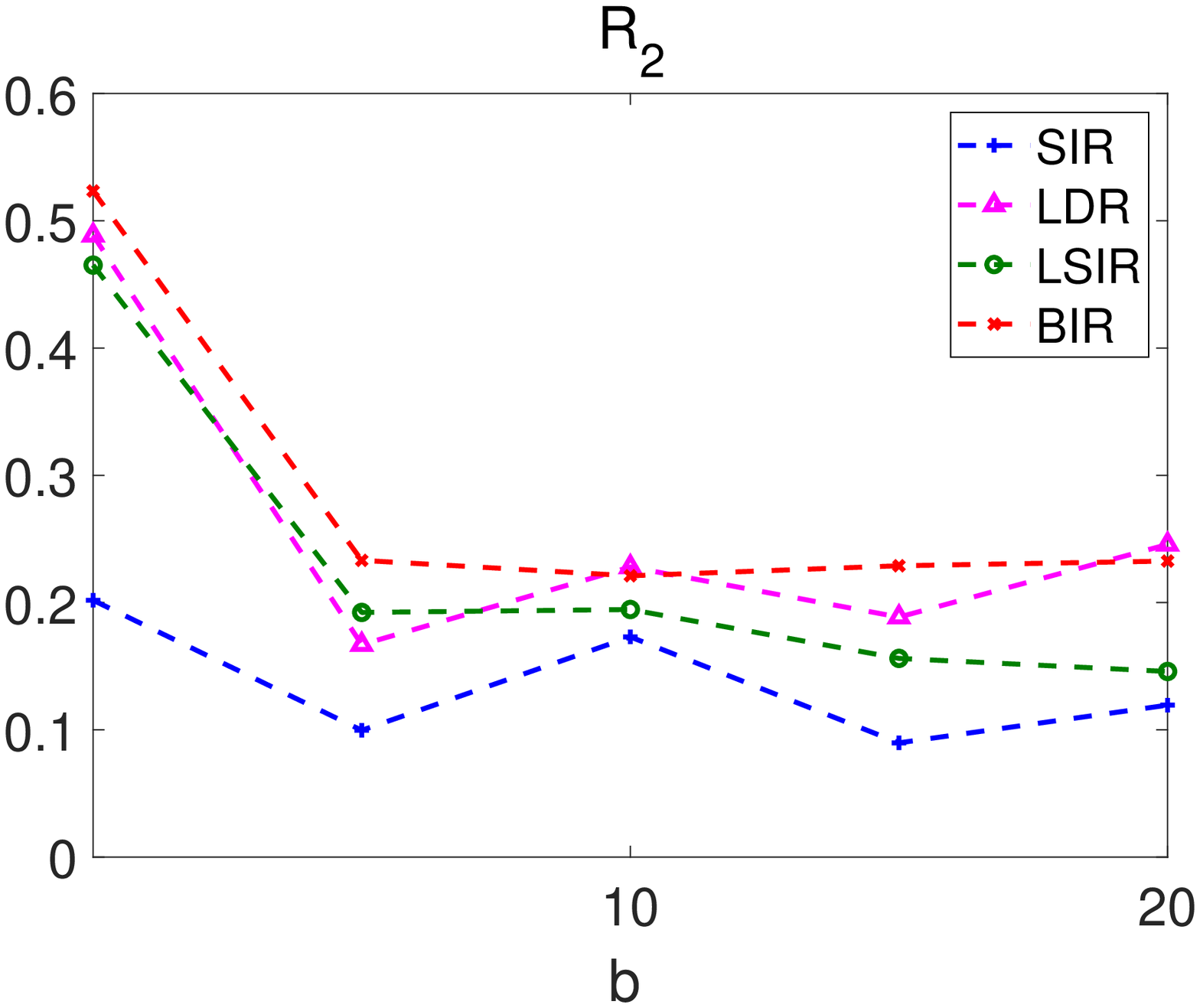}}
	\caption{The $R^2$-accuracy of the DR subspace plotted as a
	function of $b$,  for function~\eqref{e:fun3} (left) and function~\eqref{e:fun4} (right) respectively.} \label{f:eg2}
\end{figure*}

In our second example, we want to test the performance of the methods when the distribution of $\-x$ is strongly non-Gaussian.
 We assume $\-x$ is a 10-dimensional variable and the data are generated as follows.
First let $\-u=(u_1,u_2)$ follow a two-dimensional standard normal distribution. We then perform the following transform:
\begin{equation}
x_1=u_1,\quad x_2 = u_1-bu_1^2,
\end{equation}
where $b\geq0$. Here by varying parameter $b$ one can control how different the distribution of $\-x$ is from Gaussian.
Data of $y$ are generated from $\-u$, and so the transformation used to generating $\-x$ does not affect the data of $y$.
In this example we use the following two functions to generate $y$:
\begin{subequations}
\begin{align}
\quad y &= \frac{u_1}{0.5+(u_2+1.5)^2}+0.5\epsilon,\label{e:fun3}\\
\quad y & = \sin(5\pi u_1) + u_2^2 +0.1\epsilon,\label{e:fun4}
\end{align}
\end{subequations}
where $\epsilon\sim N(0,1)$.
In this test, we choose five different values of $b$: $b=0,\,5,\,10,\,15,\,20$ with sample size $n=100$, and we show the scatter plots
of the data points for all these cases in Fig.~\ref{f:rosenbrock},
where we can see that the resulting data points move apart from Gaussian as $b$ increases.
We plot the $R^2$ accuracy against the value of $b$ in Figs.~\ref{f:eg2} for both functions.
From the figures we can see that for function~\ref{e:fun3}, BIR clearly outperforms all the other methods for all the values of $b$,
and for function~\ref{e:fun4}, the BIR also has the best performance in all the cases,
with LDR being about the same at $b=10$ and $20$.
%The results demonstrate that BIR performs well for  non-Gaussian distributions.

\subsection{Mathematical example for BAVE}
We now consider a mathematical example which requires to consider the 2nd moments. 
Let $\-x$ be a 20 dimensional random variable following standard normal distribution, and  let 
$$y =  x_1^2 + 0.1\epsilon,$$ 
where noise $\epsilon \sim \mathcal{N}(0,1)$. It is easy to verify that $E(\-x|y) = 0$, which implies that the first moment based approach, i.e.,
		SIR, does not apply to this problem.
		
\begin{figure*}
  %\centering
 \centerline{ \includegraphics[width=.65\linewidth]{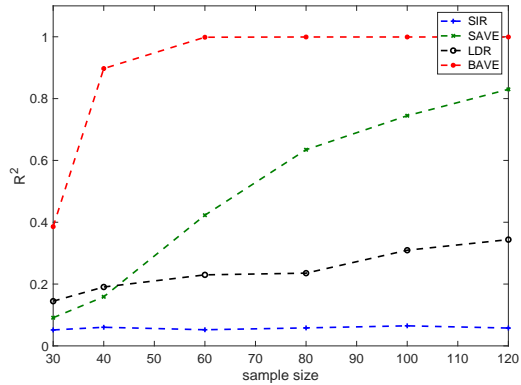}}
	\caption{The $R^2$ accuracy of the DR direction computed with different same sizes.} \label{f:bave}
\end{figure*}

		We conduct numerical experiments with six different sample sizes: 30, 40, 60, 80, 100 and 120, 
	and for each sample size,	we randomly generate 100 sets of data.
	With each set of data, we estimate the DR direction with SIR, LDR, SAVE and BAVE.
	The $R^2$ accuracy of the DR direction obtained by each method, averaged over the 100 trials, is
	shown in Fig.~\ref{f:bave}. As expected, SIR fails completely for this example -- 
	its resulting $R^2$ accuracy is near zero, regardless of the sample size. 
	The results of LDR are better than SIR but the overall accuracy remains quite low (less than $0.4$) even when the sample size
	reaches 120. 
	On the other hand,  the performance of SAVE and BAVE increases notably as the sample size increases,
	while for each sample size, the results of BAVE are considerably better than those of SAVE, 
	suggesting that BAVE performs considerably better than SAVE for this small dataset problem.

%\begin{table*}[htbp!]		
%	%\centering
%	    \begin{tabular}{lccccccccc}%{p{2cm}p{3cm}p{2cm}p{2cm}p{2cm}p{2cm}p{2cm}}
%		%{rcccccc}
%		\hline
%		\hline
%		{sample size} &  $30 $ &  $ 40 $  & $60 $ & $80$ & $100$ & $120$   \\ \hline
%		SIR   & $.0516$& $.0602$ & $.0522$ & $.0582$ & $.0649$& $.0577$\cr
%		& $\left(.0640\right )$ & $\left(.0845\right) $ & $\left(.0733\right) $& $\left(.0939\right) $ & $\left(.0898\right) $  & $\left(.0849\right)$\\ \hline
%		SAVE & $.0911$ & $.1588$ & $.4228$ & $.6346$ & $.7446$ & $.8300$ \cr
%		 & $\left(.1194\right) $      & $\left(.1793 \right) $ &$\left(.2643\right) $ & $\left(.2043\right)$& $\left(.1393\right) $ & $\left(.0732\right) $ \\\hline
%		BAVE & $ \mathbf{.3857}$ & $.\mathbf{8972}$ & $\mathbf{.9985}$ & $ \mathbf{.9993}$ & $\mathbf{ .9993}$ & $\mathbf{ .9993}$ \cr
%		& $\left(.4008\right)$ & $\left(.2625\right)$ & $\left(.0019\right)$ & $\left(4.2\times10^{-4}\right)$ & $\left(4.6e-04\right)$& $\left(5.0e-04\right)$\\
%		 \hline
%	    \end{tabular}
%			\medskip
%			
%			\caption{The mean and the standard deviation (in parenthesis) of $R^2(b)$. The best results are marked in
%			{bold}.} \label{tb:eg4}
%    \end{table*}

\subsection{Death rate prediction}
    \begin{table*}[h]
	\centering
	    \begin{tabular}{lcccccc}%{p{2cm}p{3cm}p{2cm}p{2cm}p{2cm}p{2cm}p{2cm}}
		%{rcccccc}
		\hline
		\hline
		{}{Methods} & $n = 15 $ & $n = 20 $ & $n = 25 $ & $n = 30 $  & $n = 35 $ &  $n = 40$    \\ \hline
		w/o DR & $.1832$ & $ .0855$ & $ .0551$ & $ \mathbf{.0460}$ & $ \mathbf{.0425} $ & $ \mathbf{.0380}$ \cr
		 & $\left(.2013\right) $ & $ \left(.0502\right) $ & $ \left(.0216\right)$ & $ \left( .0171\right)$ & $ \left(.0161\right) $ & $\left(.0089\right)$ \\\hline
		LDR   & -    & $.0823$      & $.0569$    & $.0490$         & $.0444$ & $.0383$      \cr
		& (-)      & $\left(.0518 \right) $ &    $\left(.0207\right) $ & $\left(.0184\right) $ & $\left(.0173\right) $ & $\left(.0089\right)$\\ \hline
		SIR   & $.4403$    & $.0982$ & $.0653$ & $.0548$ & $.0525$      & $.0430$ \cr
		& $\left(1.2417\right )$ & $\left(.0769\right) $ & $\left(.0310\right) $& $\left(.0216\right) $ & $\left(.0217\right) $  & $\left(.0108\right)$\\ \hline
		LSIR & - & $.0876$ & $.0648$ & $.0557$ & $.0485$ & $.0429$ \cr
		 & (-)       & $\left(.0461 \right) $ &    $\left(.0252\right) $ & $\left(.0224\right) $ & $\left(.0174\right) $ & $\left(.0100\right)$\\\hline
		BIR & $ \mathbf{.0484}$ & $\mathbf{.0451}$ & $\mathbf{.0465}$ & $ .0481$ & $ .0466$ & $ .0468$ \cr
		& $\left(.0110\right)$ & $\left(.0104\right)$ & $\left(.0111\right)$ & $\left(.0105\right)$ & $\left(.0114\right)$& $\left(.0126\right)$\\
		
		 \hline\hline
	    \end{tabular}	
			\medskip
			
			\caption{The mean and the standard deviation (in parenthesis) of  MRRE for Example 3. The best results are marked in
			{bold}.} \label{tb:eg3}
    \end{table*}
		    \begin{table}[h]
    	\centering
    	
    	\begin{tabular}{p{2cm}p{2cm}p{2cm}p{2cm}p{2cm}p{2cm}p{2cm}}
    		%{rcccccc}
    		\hline
    		\hline
    		Methods & min&  max\\ \hline
				w/o DR & $.0289$ & $.3374$ \\
    		LDR & $.0281$ & $.7913$ \\
    		SIR & $.0362$ & $.2534$ \\
    		LSIR & $.0380$ & $.2350$\\
    		BIR &   $\mathbf{.0247}$ &$\mathbf{ .1183}$ \\
    		 \hline\hline
    	\end{tabular}			\medskip
			
			\caption{The minimal and maximal relative regression error (RRE) in the 100 trials
			with 20 data points for the death rate example. The best results are marked in
			{bold}.} \label{tb:eg32}

    \end{table}
The example considered in this section  is to use pollution and related factors to predict the death rate~\cite{mcdonald1973instabilities,chatterjee2015regression}. This is a regression problem with 15 predictors and 60 data points and we choose this example to test how the methods perform with very small number of data. We first apply the DR  methods to select one feature (we have conducted tests with  2 and 3 features which does not improve the regression accuracy,
and so we omit those results here) and then construct a standard linear regression model of the data in the reduced dimension.
As a comparison, we also perform the regression directly without DR.
To test the methods with different numbers of  data, we perform
the experiments with $15$, $20$, $25$, $30$, $35$, $40$ data points randomly selected from the data set and
another randomly selected 20 data points used as the test set.
In each experiment we can compute the mean relative regression error (MRRE) using the data in the test set.
Specifically, suppose $\{(\-x_i,y_i)\}_{i=1}^{n_t}$ is the training set and $f_r(\cdot)$ is the regression model, the MRRE is computed as,
\[
\mathrm{MRRE} = \frac1{n_t}\sum_{i=1}^{n_t} \frac{|y_i-f_r(\-x_i)|}{y_i}.
\]
We repeat all the experiments 100 times, and  compute the mean and the standard deviation
of the obtained MRRE, which is shown in Table~\ref{tb:eg3}.
First we observe that for $n=40$ all the methods can achieve rather good accuracy;
as $n$ decrease, the results of all the other methods become evidently worse,
while that of BIR remains quite stable, suggesting that the BIR is especially effective
in the small data case.
It should be noted that for $n=15$ LDR and LSIR fail to produce reasonable results due to numerical instability,
and so we omit the results here.
More importantly it can be seen from the table that starting from $n=30$,
the regression without DR actually has the best performance,
suggesting that implementing DR is only necessary when the number of data points is below 30.
In all the cases DR is genuinely needed, i.e., $n<25$, the BIR method performs significantly better than all other methods.
To further analyze the performance, we also compute the minimal and the maximal relative regression errors (RRE)
for the 20 data-point case, and present the results in Table~\ref{tb:eg32}.
Once again, we can see that the BIR method has the best results in both the minimal and the maximal cases.
 %To  further analyze the results we show the regression plots in the reduced dimension obtained by the four methods.

  \begin{table*}[htbp!]
		
	%\centering
	    \begin{tabular}{lccccccccc}%{p{2cm}p{3cm}p{2cm}p{2cm}p{2cm}p{2cm}p{2cm}}
		%{rcccccc}
		\hline
	
		\hline
		{sample size} & $20 $ & $30 $ & $40 $ & $ 50 $  & $60 $ &  $70$ & $80$ & $90$& $100$    \\ \hline
		w/o DR & $.377$ & $ .254$ & $ .210$ & $ .203$ & $ .181 $ & $ .173$& $ .173$ & $ .169$ & $ .170$ \cr
		 & $\left(.153\right) $ & $ \left(.072\right) $ & $ \left(.035\right)$ & $ \left( .0393\right)$ & $ \left(.025\right) $ &$\left(.026\right)$ & $\left(.022\right)$ & $ \left(.022\right) $ & $\left(.022\right)$  \\\hline
		LDR   &$.394$ &$.262$& $.209$& $.198$ & $.178$ & $.173$ & $.175$&$.172$ & $.171$ \cr
		& $\left(.175\right) $      & $\left(.097 \right) $ &    $\left(.035\right) $ & $\left(.033\right) $ & $\left(.028\right) $ & $\left(.031\right)$& $\left(.024\right)$& $\left(.027\right)$& $\left(.024\right)$\\ \hline
		SIR   & $.497$& $.284$ & $.2358$ & $.217$ & $.199$& $.192$& $.193$& $.189$ & $.187$ \cr
		& $\left(.224\right )$ & $\left(.093\right) $ & $\left(.049\right) $& $\left(.044\right) $ & $\left(.035\right) $  & $\left(.034\right)$& $\left(.032\right) $ & $\left(.030\right) $  & $\left(.029\right)$\\ \hline
		LSIR & $.489$ & $.284$ & $.225$ & $.216$ & $.194$ & $.189$& $.190$ & $.183$ & $.178$ \cr
		 & $\left(.210\right) $      & $\left(.079 \right) $ &$\left(.043\right) $ & $\left(.040\right) $ & $\left(.032\right) $ & $\left(.034\right)$& $\left(.029\right) $ & $\left(.033\right) $ & $\left(.027\right)$\\\hline
		BIR & $ \mathbf{.188}$ & $.\mathbf{184}$ & $\mathbf{.178}$ & $ \mathbf{.178}$ & $\mathbf{ .167}$ & $\mathbf{ .164}$& $\mathbf{.167}$ & $ \mathbf{.165}$ & $\mathbf{.162}$ \cr
		& $\left(.034\right)$ & $\left(.034\right)$ & $\left(.031\right)$ & $\left(.029\right)$ & $\left(.025\right)$& $\left(.027\right)$& $\left(.021\right)$ & $\left(.024\right)$& $\left(.023\right)$\\
		 \hline
	    \end{tabular}
			\medskip
			
			\caption{The mean and the standard deviation (in parenthesis) of  MRRE for Example 4. The best results are marked in
			{bold}.} \label{tb:eg4}
    \end{table*}

\begin{table}[htbp!]
	\centering
	\begin{tabular}{p{2cm}p{2cm}p{2cm}p{2cm}p{2cm}p{2cm}p{2cm}}
		%{rcccccc}
		\hline
		\hline
		Methods& min&  max\\ \hline
		w/o DR  & $.1555$ & $1.273$ \\	
			LDR &$.170$ & $1.134$ \\
		SIR &$.169$& $1.874$ \\
		LSIR & $.217$ & $1.146$\\
			BIR & $\mathbf{.113}$& $\mathbf{0.287}$
		\\ \hline\hline
	\end{tabular}
	\caption{The minimal and maximal relative regression error (RRE) in the 100 trials
			with 20 data points for the automobile price example. The best results are marked in
			{bold}.}
	\label{tb:eg42}
\end{table}

\subsection{Automobile data set}
Our last example is the automobile data set in the UCI Machine Learning Repository~\cite{Dua:2019}.
The original data set contains 205 instances described by 26 attributes including 16 continuous and 10 categorical. We preprocess the data set in the following way:
we neglect the 10 categorical attributes, and  remove the instances with missing values, yielding a data set with 159 instances and 16 attributes.
We select one of the 16 attributes as the response and the others as the predictors:
specifically we want to predict the price of an automobile based the other 15 attributes of it.
In this problem we first select one feature using the DR methods, and then perform a linear regression with the selected feature.
Just like the previous example, we want to examine the performance of the DR methods in the small-data setting, i.e., a setting
where direct regression can not provide accurate results.
To do so, we conduct the experiments with $n=10,\,20,\,...,90,100$ randomly selected samples
and another 50 random samples used as the test set for all the cases.
We repeat each experiment 100 times, and compute the MRRE each time.
The mean and the standard deviation of the MRRE results are reported in Table~\ref{tb:eg4}.
From the data given in Table~\ref{tb:eg4}, we obtain rather similar conclusions  as those of Example 3.
Namely, the BIR method has the best MRRE of all the four methods used.
In Table~\ref{tb:eg42}, we show the minimal and the maximal RRE for the 20 data-point case,
and just like the results in Example~3, we find that
 the BIR method has the smallest RRE in both the minimal and the maximal cases.

\section{Conclusions}\label{sec:conclusion}

We consider dimension reduction problems for regression and we propose a Bayesian
approach for computing the conditional distribution $\pi(\-x|y)$ and perform the dimension reduction.
The method construct the likelihood function from the data with a GP regression model
and MCMC to generate samples from the conditional distribution $\pi(\-x|y)$.
%The method gets rid of the need for slicing the data points.
Numerical examples demonstrate that the proposed method is particularly effective for problems with very small data set.
We reinstate here that, due to the use of GP model, BIR does not apply to problems with very high dimensions.
Rather, we expect BIR can be useful for problems with moderately high dimensions,
and a very limited amount of data.

We believe the method can be useful in many real world applications. 
For example, in many high dimensional inverse problems and data assimilation problems,
one the data can only be informative on a small number of dimensions~\cite{cui2014likelihood,solonen2016dimension,zahm2018certified}.
A method that utilizes the DR methods to identify such data informed dimensions is currently under investigation. 
On the other hand, in certain problems gradient information is available,
and DR methods which takes advantages of the gradient information
have also been developed, e.g.~\cite{fukumizu2012gradient,constantine2015active,lam2018multifidelity}.
In this case, we expect that the gradient information can also be used to enhance the performance of the BIR method,
via, for example, Gradient-Enhanced Kriging~\cite{morris1993bayesian}, and we plan to investigate this problem in the future.

\bibliographystyle{plain}
\bibliography{bir}

\end{document}